\documentclass[sigconf,authorversion, review=false]{acmart}

\usepackage{array}
\usepackage{multirow}
\usepackage{pifont}% http://ctan.org/pkg/pifont
\usepackage{listings}
\usepackage[textwidth=0.7cm,colorinlistoftodos]{todonotes}
\usepackage{diagbox}
\usepackage{color}
\usepackage{soul}
\usepackage{tabularx}
\usepackage{multirow}
\usepackage{subcaption}
\usepackage{xcolor}

\definecolor{darkgreen}{RGB}{0,127,0}
\definecolor{darkred}{RGB}{127,0,0}
\definecolor{shadecolor}{RGB}{250, 250, 250}
\definecolor{lightyellow}{RGB}{250, 250, 212}
\definecolor{HLYELLOW}{RGB}{240, 127, 0}
\definecolor{hlyellow}{RGB}{240, 127, 0}
\sethlcolor{white}
\graphicspath{ {./img/} }
\definecolor{peyellow}{RGB}{240, 253, 255}
\definecolor{mycyan}{RGB}{159, 207, 255}
\sethlcolor{mycyan}

\usepackage{soul}
\soulregister\cite7
\soulregister\ref7
\soulregister\pageref7
\soulregister\lstinline7

\usepackage[normalem]{ulem} % use normalem to protect \emph

\definecolor{revisionhighlight}{rgb}{0.8, 0.8, 1}
\DeclareRobustCommand{\revision}[1]{#1}

\definecolor{sfcolor1}{HTML}{1b9e77}
\definecolor{sfcolor2}{HTML}{d95f02}
\definecolor{sfcolor3}{HTML}{e6ab02}
\definecolor{sfcolor4}{HTML}{e7298a}
\definecolor{sfcolor5}{HTML}{7570b3}
\colorlet{hightlightcolor1}{sfcolor1!30!white}
\colorlet{hightlightcolor2}{orange!20!white}
\colorlet{hightlightcolor3}{yellow!30!white}
\colorlet{hightlightcolor4}{sfcolor4!20!white}
\colorlet{hightlightcolor5}{cyan!20!white}

\DeclareRobustCommand{\revisiona}[1]{#1}
\DeclareRobustCommand{\revisionb}[1]{#1}
\DeclareRobustCommand{\revisionc}[1]{#1}
\DeclareRobustCommand{\revisiond}[1]{#1}
\DeclareRobustCommand{\revisione}[1]{#1}

\DeclareRobustCommand{\revisiondnotag}[1]{#1}

\DeclareRobustCommand{\revision}[2]{#2}

\usepackage{tikz}
\usetikzlibrary{positioning,fit,calc,shapes.geometric}

\AtBeginDocument{%
  \providecommand\BibTeX{{%
    \normalfont B\kern-0.5em{\scshape i\kern-0.25em b}\kern-0.8em\TeX}}}

\setcopyright{acmcopyright}
\copyrightyear{2021}
\acmYear{2021}
\acmDOI{10.1145/1122445.1122456}

\acmConference[SC'21]{Supercomputing '21: The International Conference for High Performance Computing, Networking, Storage, and Analysis}{November 14--19, 2021}{St. Louis, MO}
\acmBooktitle{Supercomputing '21: The International Conference for High Performance Computing, Networking, Storage, and Analysis,
  November 14--19, 2021, St. Louis, MO}
\acmPrice{15.00}
\acmISBN{978-1-4503-XXXX-X/18/06}

\begin{document}

\title[Productivity, Portability, Performance: Data-Centric Python]{Productivity, Portability, Performance: Data-Centric Python}

\author{Alexandros Nikolaos Ziogas, Timo Schneider, Tal Ben-Nun, Alexandru Calotoiu, Tiziano De Matteis, Johannes de Fine Licht, Luca Lavarini, and Torsten Hoefler}
\affiliation{
    \institution{Department of Computer Science, ETH Zurich}
    \country{Switzerland}
}

\renewcommand{\shortauthors}{Ziogas et al.}

\begin{abstract}
    Python has become the \textit{de facto} language for scientific computing.
    Programming in Python is highly productive, mainly due to its rich science-oriented software ecosystem built around the NumPy module.
    As a result, the demand for Python support in High Performance Computing (HPC) has skyrocketed.
    However, the Python language itself does not necessarily offer high performance.
    In this work, we present a workflow that retains Python's high productivity while achieving portable performance across different architectures.
    The workflow's key features are HPC-oriented language extensions and a set of automatic optimizations powered by a data-centric intermediate representation.
    We show performance results and scaling across CPU, GPU, FPGA, and the Piz Daint supercomputer (up to 23,328 cores), with 2.47x and 3.75x speedups over previous-best solutions, first-ever Xilinx and Intel FPGA results of annotated Python, and up to 93.16\% scaling efficiency on 512 nodes.
\end{abstract}

\maketitle

\section{Introduction}
\label{sec:intro}

Python is \textit{the} language to write scientific code~\cite{octoverse}.
The capability to write and maintain Python code with ease, coupled with a vast number of domain-specific frameworks and libraries such as SciPy, Matplotlib, scikit-learn~\cite{sklearn}, or pandas~\cite{pandas}, leads to high productivity.
It also promotes collaboration with reproducible scientific workflows shared using Jupyter notebooks~\cite{jupyter}.
Therefore, numerous scientific fields, ranging from machine learning~\cite{pytorch, tensorflow} to climate~\cite{gt4py} and quantum transport~\cite{dace-omen} have already adopted Python as their language of choice for new developments.

\begin{figure}[t]
    \centering
    \includegraphics[width=.95\linewidth]{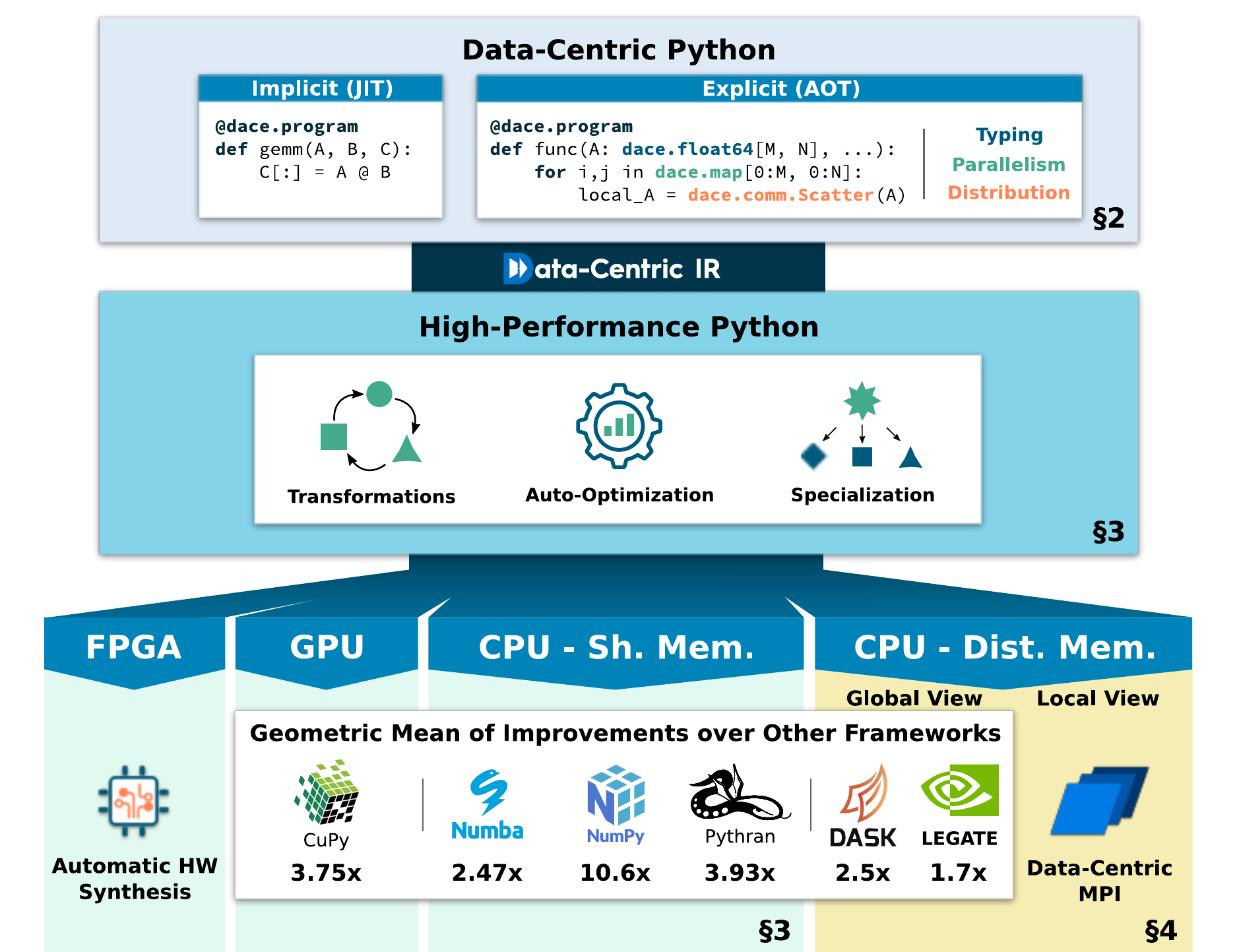}
    \vspace{-1em}
    \caption{Data-Centric Python overview.}
    \vspace{-1em}
    \label{fig:overview}
\end{figure}

As a result, the scientific community now pushes to also make Python \textit{the}
language for writing high-performance code.
For most scientific applications, NumPy arrays~\cite{numpy} provide the core data structures, interfaces, and BLAS and LAPACK library interoperability.
NumPy is optimized to provide efficient data structures and fast library implementations for many common operations.
However, the performance benefits of NumPy are tied to optimized method calls and vectorized array operations, both of which evaporate in larger scientific codes that do not adhere to these constraints.
Therefore, there is a significant performance gap between the \textit{numpythonic} code-style and general code written by domain scientists.

In HPC, the three \textbf{P}s (\textbf{Productivity, Portability, Performance}) are driving recent developments in infrastructure and programming model research to ensure sustainability~\cite{productivity1,productivity2}.
In Python, this drive has resulted in many optimized domain-specific libraries and frameworks~\cite{tensorflow,pytorch,jax,gt4py,devito}.
Simultaneously, the diversity of the hardware landscape motivated the creation of interface libraries, such as CuPy~\cite{cupy}, which provides replacements to NumPy operations for NVIDIA and AMD GPUs, and MPI4PY~\cite{mpi4py}, which offers direct MPI bindings.
Lower-level interfaces, such as Cython~\cite{cython}, promise high performance at the cost of writing code that resembles C, and lazy evaluation of array operations~\cite{bohrium,legate,pytorch} that enable high-performance runtime systems.
Furthermore, a variety of JIT compilers~\cite{numba,pythran,jax} address the performance degradation resulting from the interpreter.
Last but not least, runtime systems~\cite{legate}, distributed tasking (Dask)~\cite{dask}, and remote procedure calls~\cite{ray} further scale Python to distributed systems.
Despite the abundance of choices, Python still struggles: while each approach works towards one or more of the \textbf{P}s, none of them supports all at the same time.

We propose a way to bridge the gap between the three \textbf{P}s for Python programming using a data-centric paradigm. In particular, we empower Python users with an automatic optimization and specialization toolbox, which spans the entire Python/HPC ecosystem (Fig.~\ref{fig:overview}) --- from the code, through communication distribution, to hardware mapping.
At the core of the toolbox, we use the Stateful Dataflow multiGraphs (SDFG)~\cite{dace}  data-centric intermediate representation, which enables these optimizations in the form of multi-level data movement transformations. 
With a data-centric representation, as opposed to library bindings, all data dependencies and potential overlap are inferred statically from the code, and interpreter overhead is mitigated. 
Compared with implicit and lazy evaluation approaches, we also provide a set of extensions that give power users complete control over parallelism and partitioning schemes, using \textit{pythonic} principles and syntax (e.g., returning ``local view'' objects from global data, but allowing users to operate on the ``global view'' as well).

We demonstrate a wide variety of benchmarks using the automatic toolbox \textit{\textbf{over annotated Python code}}, both on individual nodes and the Piz Daint supercomputer.
For the former, we show that it consistently outperforms other automatic approaches on multicore CPUs and GPUs, and \textit{for the first time} show automatic Python HPC compilation results for both Xilinx and Intel FPGAs, which vastly differ in architecture and programming language.
In distributed-memory environments, we show high scaling efficiency and absolute performance compared with distributed tasking.
Thus, we realize all three \textbf{P}s within a single system.

The paper makes the following contributions:
\begin{itemize}
    \item (\textbf{Productivity}) Definition of high-performance Python, a methodology to translate it to a data-centric IR, and extensions to improve said conversion via explicit annotation.
    \item (\textbf{Portability}) A set of automatic optimizations for CPU, GPU and FPGA, outperforming the best prior approaches by 2.47$\times$ on CPU and 3.75$\times$ on GPU on average (geometric mean~\cite{geomean}).
    \item (\textbf{Performance}) Automatic implicit MPI transformations and communication optimizations, as well as explicit distribution management, with the former scaling to 512 nodes with up to 93.16\% efficiency.
\end{itemize}

%. 

\section{Data-centric Python}

The central tenet of our approach is that understanding and optimizing data movement is the key to portable, high-performance code.
In a data-centric programming paradigm, three governing principles guide development and execution: 
\begin{enumerate}
    \item Data containers must be separate from computations.
    \item Data movement must be explicit, both from data containers to computations and to other data containers.
    \item Control flow dependencies must be minimized, they shall only define execution order if no implicit dataflow is given.
\end{enumerate}

\revisione{In the context of SDFGs, examples of data containers are arrays and scalar data, which have a NumPy-compatible data type, such as \texttt{int32} or \texttt{float64}.}

Python is an imperative language and, therefore, not designed to express data movement.
Its terseness makes the process of understanding dataflow difficult, even when comparing to other languages like C and FORTRAN, as the types of variables in Python code cannot be statically deduced. 

Below, we define high-performance Python programs, discuss the decorators that we must add to Python code to make the dataflow analyzable, and then detail how we translate them into the SDFG data-centric intermediate representation.

\subsection{High Performance Python}

Our approach supports a large subset of the Python language that is important for HPC applications. 
The focus lies on NumPy arrays~\cite{numpy} and operations on such arrays.
In addition to the low-overhead data structures NumPy offers, it is central to many frameworks focused on scientific computing, e.g., SciPy, pandas, Matplotlib.
As opposed to lazy evaluation approaches, high-performance Python must take control flow into account to auto-parallelize and avoid interpreter overhead.
This tradeoff between performance and productivity is necessary because Python features such as co-routines are not statically analyzable and have to be parsed as ``black-boxes''.
To combat some of these Python quirks, we propose to augment the language with analyzable constructs useful for HPC.

\subsection{Annotating Python}
\label{sec:augmenting}
The Data-Centric (DaCe) Python frontend parses Python code and converts it to SDFGs on a per-function basis.
The frontend will parse only the Python functions that have been annotated explicitly by the user with the \lstinline{@dace.program} decorator. DaCe programs can then be called like any Python function and perform Just-in-Time (JIT) compilation.

\paragraph{Static symbolic typing}
To enable Ahead-of-Time (AOT) compilation, which is of key importance for FPGAs and for reusing programs across different inputs, SDFGs should be statically typed.
Therefore, the function argument data types are given as type annotations, providing the required information as shown below:
\begin{lstlisting}
N = dace.symbol()
@dace.program
def jacobi_1d(TSTEPS: dace.int32,
              A: dace.float64[N],
              B: dace.float64[N]):
    
    for t in range(1, TSTEPS):
        B[1:-1] = 0.33333 * (A[:-2]+A[1:-1]+A[2:])
        A[1:-1] = 0.33333 * (B[:-2]+B[1:-1]+B[2:])
\end{lstlisting}
The Python method \lstinline{jacobi_1d} has three arguments; \lstinline{TSTEPS} is a 32-bit integer scalar, while \lstinline{A} and \lstinline{B} are double precision floating-point vectors of length \lstinline{N}.
The symbolic size \lstinline{N}, defined with \lstinline{dace.symbol}, indicates that the vector sizes can be dynamic (but equal). All subsets are then symbolically defined (e.g., the subset \lstinline{B[1:-1]} becomes \lstinline{B[1:N-1]}, and symbolic manipulation can then be performed in subsequent data-centric transformations.

\paragraph{Parametric parallelism}
An important feature that has no direct expression in Python is a loop that can run in parallel.
Our approach supports explicit parallelism declaration through map scopes, similarly to an N-dimensional \texttt{parallel for}.
There are two ways to take advantage of this feature.
DaCe provides the \lstinline{dace.map} iterator, which can be used in Python code as a substitute to the Python built-in \lstinline{range} iterator and generates a map scope when parsed:
\begin{lstlisting}
for i, j in dace.map[0:M, 0:N]:
    A[i, j] = B[j, i]
\end{lstlisting}
Alternatively, the DaCe framework provides a \lstinline{LoopToMap} transformation that detects for-loops in the IR, whose iterations can be executed safely in parallel (using symbolic affine expression analysis), and converts them to map scopes automatically.

\subsection{From Python to DaCe}
\label{sec:py2dc}

\begin{table}[t]
\small
\begin{tabular}{p{1.4in} p{1.6in}}
\toprule
\bf Python &	\bf SDFG Equivalent \\
\midrule
\multicolumn{2}{l}{\bf Declarations and Types}\\
\midrule
Primitive data types & Scalar data container\\\addlinespace
NumPy array &Array data container\\\addlinespace
\midrule
\multicolumn{2}{l}{\bf Assignments }\\
\midrule
Assignments & Tasklet or map scope with incoming and outgoing memlets for read/written operands  \\\addlinespace
Array subscript & Memlet (dataflow edge) \\
\midrule
\multicolumn{2}{l}{\bf Statements}\\
\midrule
Branching (\texttt{if})&	Branch conditions on state transition edges\\\addlinespace Iteration (\texttt{for})&	Conditions, increments on state transition edges \\\addlinespace
Control-flow \hspace{1in} (\texttt{break, continue, return})	& Edge to control structure or function exit state\\\addlinespace

\midrule
\multicolumn{2}{l}{\bf Functions  }\\
\midrule
Function calls (with source) and decorator for argument types&	Nested SDFG for content, memlets reduce shape of inputs and outputs\\\addlinespace
External/Library calls &	Tasklet with callback or Library Node
\\\addlinespace
\midrule
\end{tabular}
\caption{Mapping of Python syntax and constructs to SDFG.}\vspace{-2em}
\label{table:py2dace}
\end{table}

\begin{figure}[t]
    \centering
    \begin{subfigure}{0.235\textwidth}
        \includegraphics[page=1,width=\textwidth]{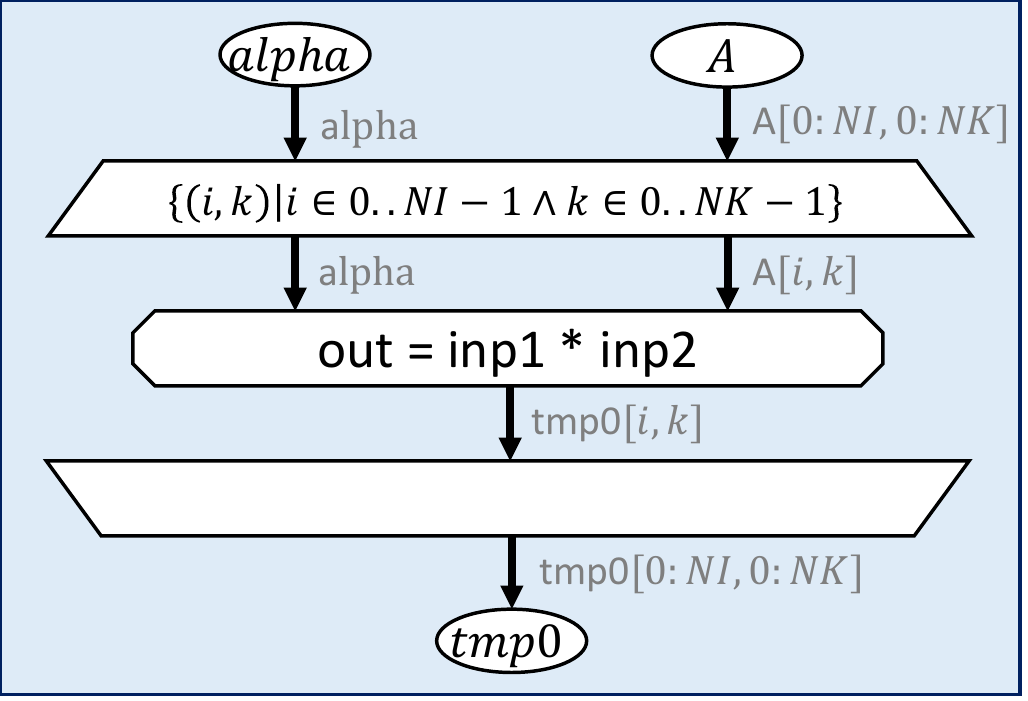}
        \vspace{-1em}
        \caption{Element-wise array operation \lstinline{tmp0 = alpha * A}.}
        \label{fig:sdfg-repr-elwiseop}
    \end{subfigure}
    \begin{subfigure}{0.235\textwidth}
        \includegraphics[page=3,width=\textwidth]{img/DaCe_P4_SC21.pdf}
        \vspace{-1em}
        \caption{Augmented assignment with WCR.}
        \label{fig:sdfg-repr-augassign-wcr}
    \end{subfigure}
    \caption{SDFG representations}
\end{figure}

We turn to present the SDFG intermediate representation (IR) and a novel data-centric Python translation procedure in tandem.
While previous work~\cite{dace} converted a restricted, low-level Python definition of the SDFG IR, here we aim to cover the majority of the Python/NumPy language constructs via static analysis and fallback for unsupported features.
We summarize the equivalence between Python constructs and SDFG counterparts in Table~\ref{table:py2dace}, and present the generation of an SDFG from a Python program using the \lstinline{gemm} kernel as an example:
\begin{lstlisting}
@dace.program
def gemm(alpha, beta, C, A, B):
    C[:] = alpha * A @ B + beta * C
\end{lstlisting}
Our first pass traverses the Python AST to simplify the expressions into individual steps, similar to static single assignment~\cite{ssa}, respecting order of operations:

\begin{lstlisting}
    tmp0 = alpha * A
    tmp1 = tmp0 @ B
    tmp2 = beta * C
    C = tmp1 + tmp2
\end{lstlisting}

The first step in the above code multiplies each element of \lstinline{A} with \lstinline{alpha}.
SDFGs view data containers separately from the computations the data are part of, as per the first data-centric tenet.
These containers are represented by oval \textit{Access nodes}.
In the first statement, these refer to \lstinline{tmp0}, \lstinline{alpha}, and \lstinline{A} (see Fig.~\ref{fig:sdfg-repr-elwiseop}).

In SDFGs, connections to data containers are called \textit{memlets}, and they describe the data movement --- the edge direction indicates whether it is read or written, and its contents refer to the part of the data container that is accessed.
Computations consume/produce memlets and can be divided into multiple types:
\begin{enumerate}
    \item Stateless computations (\textit{Tasklets}, shown as octagons), e.g., representing scalar assignments such as \lstinline{a = 1}.
    \item Calls to external libraries (\textit{Library Nodes}, folded rectangles), that represent calls to functions that are not in the list of functions decorated with \lstinline{dace.program}. Matrix-matrix multiply is a common and important operation \lstinline{tmp1 = tmp0 @ B} and is contained in a Library Node called \textit{MatMul}.
    \item Calls to other SDFGs (\textit{Nested SDFGs}, rectangles), which represent calls to functions decorated with \lstinline{dace.program}.
    \item \textit{Maps} are a particular type of Nested SDFGs matching the language augmentation previously discussed in Section~\ref{sec:augmenting} and express that the content can be processed in parallel.
\end{enumerate}

Element-wise array operations automatically yield Map scopes. Similarly, assignments to arrays yield a Map scope, containing a Tasklet with the element-wise assignment. Augmented assignments such as \lstinline{C += 1} are a special case where the output is also an input when no data races are detected.

Parallel maps can be augmented to express how \textit{Write-Conflict Resolution} (WCR) should determine the value of data when multiple sources write to it concurrently.
If data races are found, the outgoing edges are marked as dashed.
E.g., the following program requires WCR (SDFG representation is shown in Fig.~\ref{fig:sdfg-repr-augassign-wcr}):
\begin{lstlisting}
for i, j in dace.map[0:NI, 0:NJ]:
    alpha += C[i, j]
\end{lstlisting}

By connecting the inputs and outputs of computations with the containers explicitly, the data-centric representation of a program can be created. To represent control dependencies, we encapsulate code driven by pure data dependencies in \textit{States} (bright blue regions) in the SDFG. The states are connected with \textit{State Transition Edges} (in blue) that can be annotated with control flow, representing loops, branch conditions, and state machines in general. The following Python program is represented by the SDFG in Fig.~\ref{fig:sdfg-repr-for-loop}:
\begin{lstlisting}
for i in range(NI):
    C[i] += 1
\end{lstlisting}
\begin{figure}[h]
    \centering
    \includegraphics[page=4,width=.9\columnwidth]{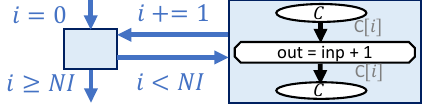}
    \vspace{-1em}
    \caption{SDFG representation of a Python for-loop. There are two states (left: guard, right: body) connected by control flow (state transition) edges that define the loop range.}
    \label{fig:sdfg-repr-for-loop}
\end{figure}
\revisiona{The above loop also is an excellent use case for the \lstinline{LoopToMap} transformation (Section~\ref{sec:augmenting}) since its iterations are independent.}

In the conversion to the SDFG IR, we also replace calls to library functions (e.g., \lstinline{np.linalg.solve}) and object methods (\lstinline{A.view()}) with custom subgraphs or Library Nodes, which users can extend for other libraries and object types via a decorated function. Following this initial conversion, the resulting SDFG contains a state per statement and a nested SDFG per function call.

\subsection{Dataflow Optimization} \label{sec:dfopt}

The direct translation to SDFGs creates a control-centric version of the code, which matches the Python semantics but does not contain dataflow beyond a single statement (similarly to compilers' \texttt{-O0}). To mitigate this, we run a pass of IR graph transformations that coarsens the dataflow, exposing a true data-centric view of the code (similar to \texttt{-O1}). The transformations include redundant copy removals, inlining Nested SDFGs, and others (14 in total), which only modify or remove elements in the graph, such that they cannot be applied indefinitely.

\begin{figure}[t]
    \centering
    \includegraphics[page=5,width=\columnwidth]{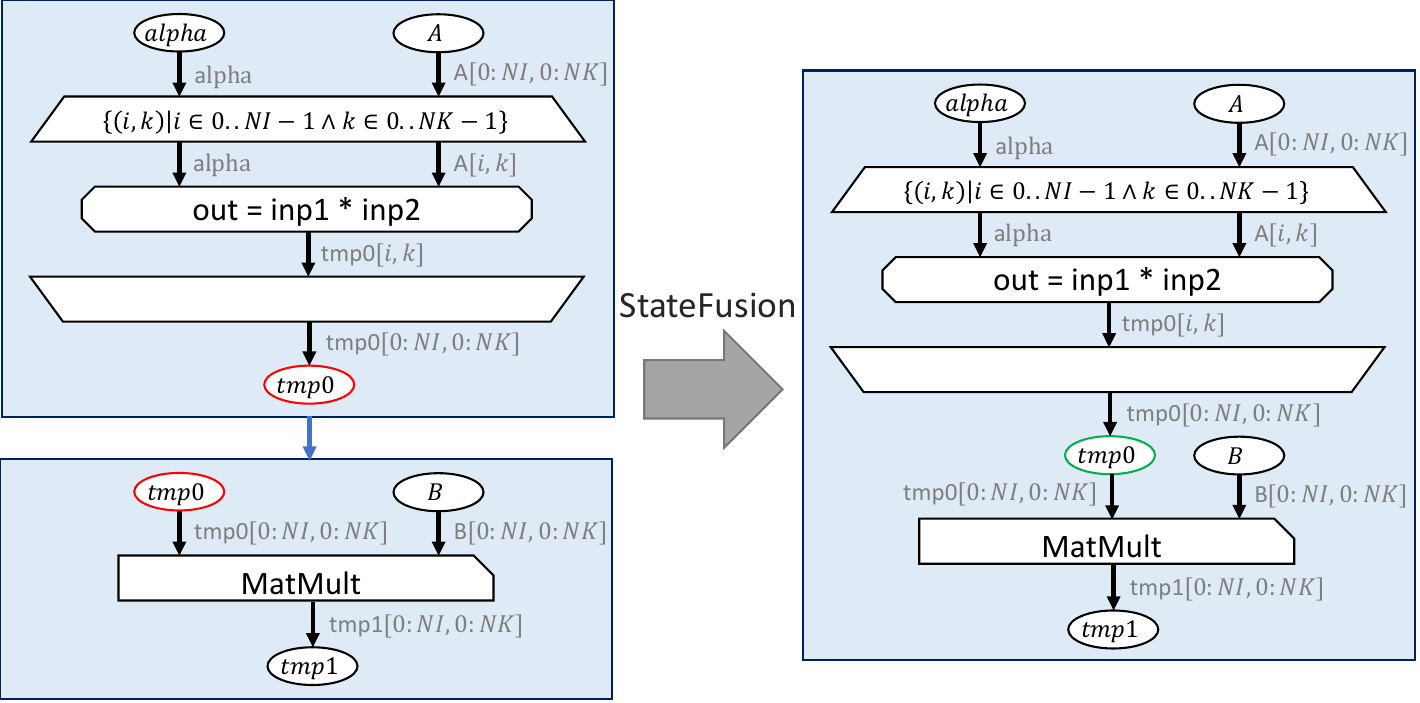}
    \vspace{-1em}
    \caption{State fusion of \lstinline{tmp0 = alpha * A} and \lstinline{tmp1 = tmp0 @ B}.}
    \label{fig:sdfg-repr-statefusion}
\end{figure}

To understand this pass, we showcase one transformation --- state fusion --- which allows merging two states where the result does not produce data races. 
For example, the two states containing the assignments below can be merged: 
\begin{lstlisting}
tmp0 = alpha * A
tmp1 = tmp0 @ B
\end{lstlisting}
Internally, the transformation matches a subgraph pattern of two states connected together and compares source and sink Access nodes using symbolic set intersection. If no data dependency constraints are violated, Access nodes are either fused (if they point to the same memory, see Fig.~\ref{fig:sdfg-repr-statefusion}) or set side by side, creating multiple connected components that can run in parallel.

All transformations in the DaCe framework follow the same infrastructure, either matching a subgraph pattern or allowing the user to choose an arbitrary subgraph~\cite{dace}. While the dataflow coarsening pass happens automatically as part of our proposed toolbox, one can also apply transformations manually and separately, without changing the original Python source code (we color such ``performance engineering codes'' in cyan):
\begin{lstlisting}[backgroundcolor=\color{peyellow}]
sdfg = gemm.to_sdfg()
sdfg.apply(StateFusion)
\end{lstlisting}

\begin{figure*}[t]
    \centering
    \includegraphics[page=10, width=\textwidth]{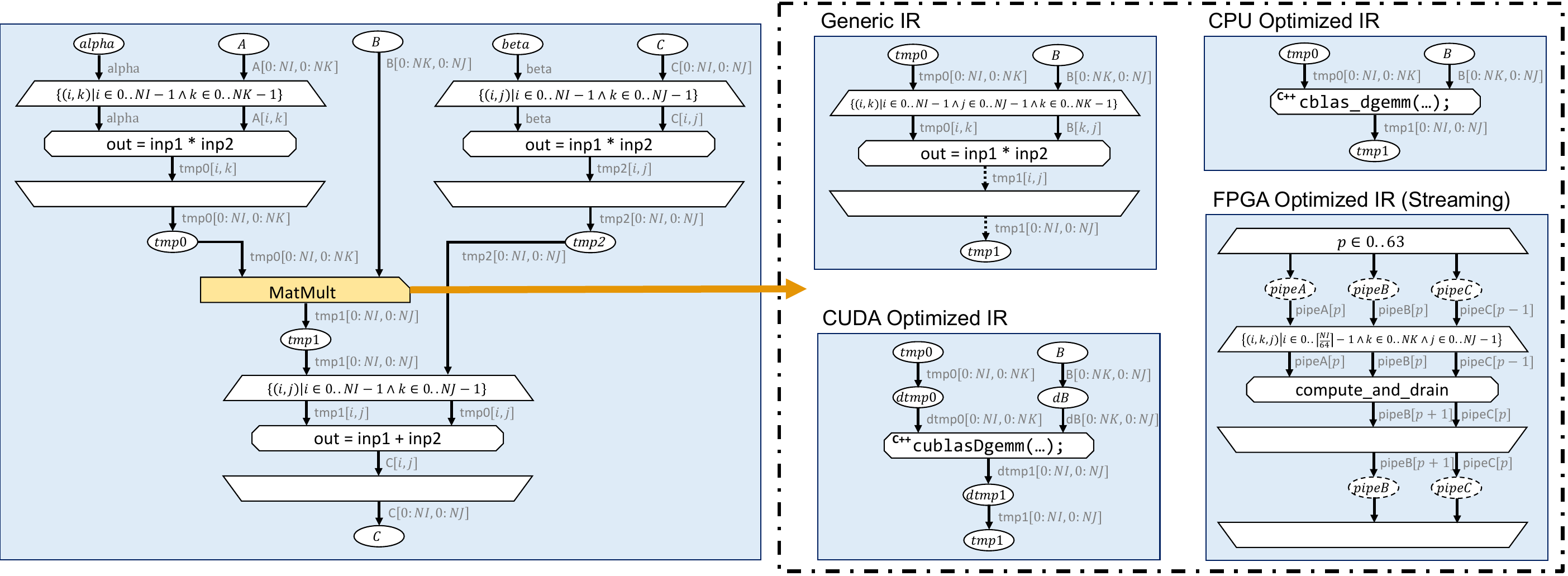}
    \vspace{-1em}
    \caption{Dataflow-coarsened GEMM SDFG and specializations for different architectures.}
    \label{fig:gemm-strict}
\end{figure*}

\subsection{Python Restrictions}

Some features available in Python are incompatible with our definition of high performance Python, and are discussed below.
This does not exclude programs using the full feature set of Python from analysis, but calls to functions containing unsupported features will not benefit from our optimization.
The restricted features are:
\begin{enumerate}
    \item Python containers (lists, sets, dictionaries, etc.) other than NumPy arrays as arguments, as they are represented by linked lists and difficult to analyze from a dataflow view. Note that this does not preclude internal containers (which can be analyzed) and list comprehensions.
    \item Dynamic typing: fixed \lstinline{struct}s are allowed in SDFGs, so fields can be transformed and their class methods into functions decorated with the \lstinline{dace.program} decorator. However, dynamic changes to classes or dynamic types are unsupported as their structure cannot be statically derived. 
    \item Control-dependent variable state (i.e., no scoping rules), e.g., the following valid Python:
\begin{lstlisting}
x = ...
if x > 5:
    y = np.ndarray([5, 6], dtype=np.float32)
# use y (will raise exception if x <= 5)
\end{lstlisting}
        \item Recursion. This is a limitation of the data-centric programming model~\cite{dace}, as recursion is a control-centric concept that is not portable across platforms.
\end{enumerate}

After performing the full translation and coarsening, the resulting \lstinline{gemm} kernel can be seen in Figure~\ref{fig:gemm-strict} (left-hand side). This data-centric representation can now use the SDFG IR capabilities to further optimize and map the original Python code to different architectures and distributed systems.

\section{Portability and Performance}
\label{sec:portable}

The translated data-centric Python programs can now be optimized for performance on different hardware architectures. In this section, we propose a novel set of data-centric passes to auto-optimize and specialize SDFGs to run at state-of-the-art performance on CPUs, GPUs, and FPGAs, all from the same source IR.
\revisionc{The programming portability is very high since our approach starts from the same Python codes parsed to the same SDFGs. Although the final optimized IRs may differ, the process can be automatic, and the user needs only to select the architecture to specialize for. In our evaluation, we only discuss results produced in an \textbf{automated} fashion.}

\subsection{Automatic Optimization Heuristics}
\label{sec:autoopt}

As mentioned above, DaCe provides a user-extensible set of graph transformations. Yet, the framework does not perform them automatically~\cite{dace}, to endow performance engineers with fine-grained control and promote separation of concerns.
\revisiona{Furthermore, DaCe includes tools and graphical interfaces to assist users with manual optimization without the explicit need for an expert.}
For productivity purposes, however, we believe that prototyping fast data-centric Python programs \textit{should} be possible with minor code modifications.

By observing the common pitfalls in generated code from SDFGs vs. what a performance engineer would write, we propose a set of transformation heuristics for SDFGs that yield reasonable performance in most cases (\texttt{-O3} compiler equivalent).
This pass can be performed automatically (configurable) or using the following decorator:
\begin{lstlisting}
@dace.program(auto_optimize=True, device=...)
\end{lstlisting}
where \lstinline{device} can be \lstinline|DeviceType.{CPU,GPU,FPGA}|.

Our auto-optimizer performs the following passes in order:
\begin{enumerate}
    \item \textbf{Map scope cleanup}: Remove ``degenerate'' maps of size 1, repeatedly apply the \textit{LoopToMap} transformation (Section \ref{sec:augmenting}), and collapse nested maps together to form multidimensional maps. The latter also increases the parallelism of GPU kernels as a by-product.
    \item \textbf{Greedy subgraph fusion}: Collect all the maps in each state, fusing together the largest contiguous subgraphs that share the same (or permuted) iteration space or the largest subset thereof (e.g., fusing the common three dimensions out of four). We use symbolic set checks on memlets to ensure that the data consumed is a subset of the data produced.
    \item \textbf{Tile WCR maps}: Tile (configurable size) parallel maps with write-conflicts that result in atomics, in order to drastically reduce such operations.
    \item \textbf{Transient allocation mitigation}: Move constant-sized and small arrays to the stack, and make temporary data containers persistent (i.e., allocated upon SDFG initialization) if their size only depends on input parameters. This nearly eliminates dynamic memory allocation overhead.
\end{enumerate}

Beyond the above general-purpose heuristics, we apply more transformations depending on the chosen device: For CPUs, we try to increase parallelism by introducing the OpenMP \lstinline{collapse} clause. For GPU and FPGA, we perform the \lstinline|{GPU,FPGA}TransformSDFG| automatic transformations~\cite{dace}, which introduce copies to/from the accelerator and convert maps to accelerated procedures. 

On the FPGA, we perform a few further transformations that diverge from the ``traditional'' fused codes in HPC: we create separate connected components (regions on the circuit) to stream off-chip (DRAM) memory in bursts to the program. Between computations, we try to modify the graph's structure to be composed of separate pipelined units that stream memory through FIFO queue Access nodes (we call this transformation \lstinline{StreamingComposition}). This also enables further transformations to the graph to create systolic arrays during hardware specialization. 

From this point, the only remaining step to lower the SDFG is to specialize the Library Nodes to their respective fastest implementations based on the target platform.

\subsection{Library Specialization}
\label{sec:portability}

Library Nodes, such as the \lstinline{MatMul} operation in \lstinline{gemm}, can be expanded to a wide variety of implementations. An \textit{expansion} is defined similarly to a transformation, as a replacement subgraph to a single node, and can specialize its behavior. 

We demonstrate specializations of matrix multiplication in Fig.~\ref{fig:gemm-strict}. In particular, one can expand a Library Node into a C++ tasklet, calling an external function from MKL or CUBLAS (CPU, CUDA optimized IR in the figure); into an optimized subgraph (e.g., FPGA optimized IR, here using the Streamed expansion); or into a ``native'' SDFG subgraph (Generic IR).

In the automatic heuristics, we employ a priority list of implementations per platform, starting from the fast library calls, through optimized versions, and if all fail to expand (e.g., if the multiplication occurs within a kernel), we expand to the ``native'' SDFG. 
This yields an SDFG that can both be further optimized by performance engineers, and execute at reasonable performance out-of-the-box.

\revisionb{Users can employ the DaCe API to create their own libraries and Library Nodes~\cite{licht2021stencilflow}. This process is comparable to creating bindings.}

\subsection{Ahead-of-Time Compilation}
\revisiona{
DaCe provides extensive support for AOT compilation, due to the risk of overheads JIT compilation introduces in an HPC environment.
Namely, DaCe provides the capability of AOT compilation if the decorated function's arguments are type-annotated as described in Section~\ref{sec:augmenting}.
A decorated function or an SDFG can be directly compiled to a shared library through a Python script.
Alternatively, the user can employ the \textit{sdfgcc} command-line tool to compile SDFGs.}

\revisione{The compilation process utilizes specialized backends to generate optimized code for each supported architecture~\cite{dace}.
For example, C++ code is generated for CPUs, while the Nvidia GPU backend outputs C++ and CUDA programs.
For FPGAs we utilize Vitis HLS and OpenCL for Xilinx and Intel architectures, respectively.}

\begin{figure}[t]
    \centering
    \includegraphics[width=.95\columnwidth]{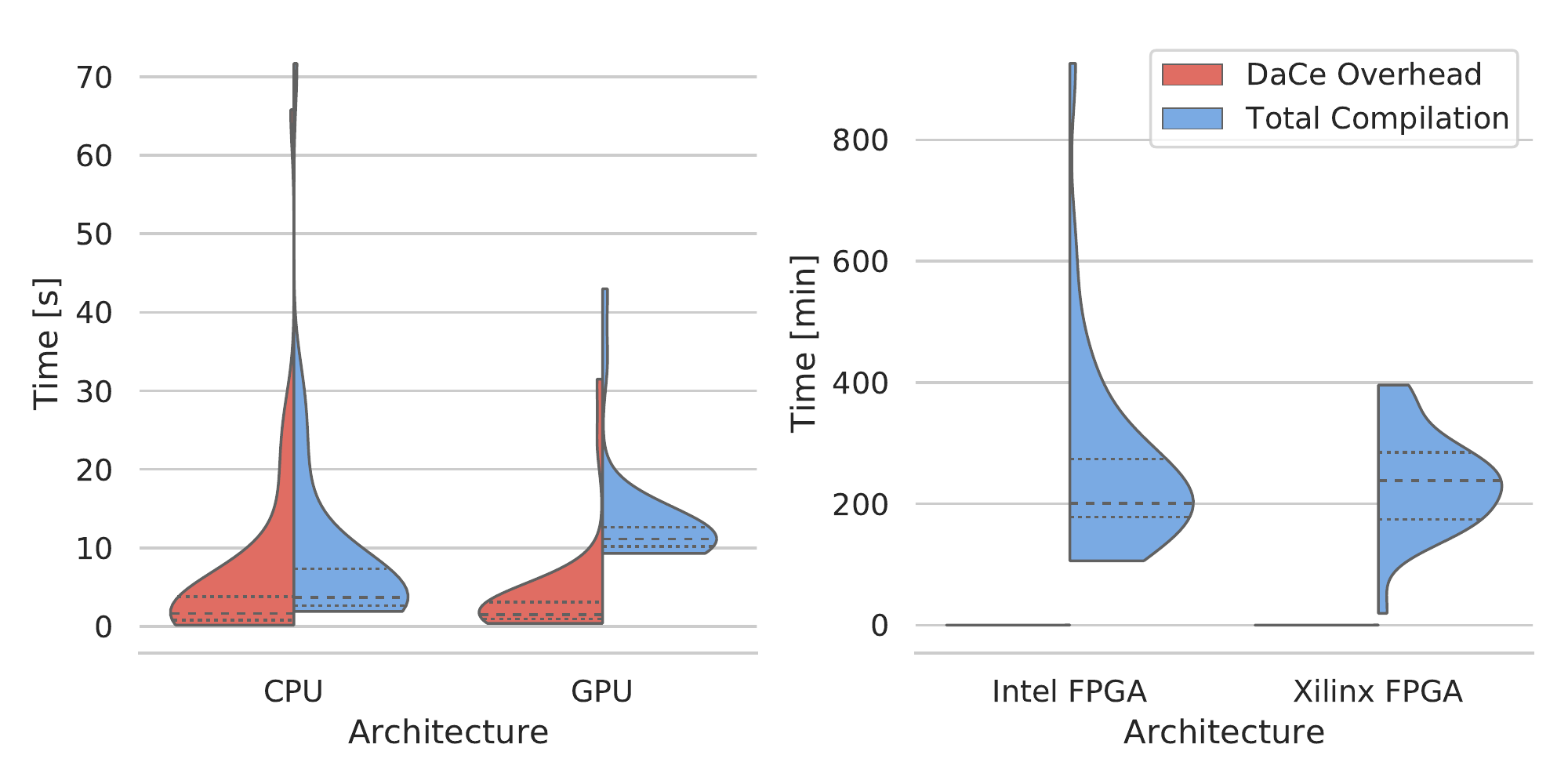}
    \vspace{-1em}
    \caption{Distributions of DaCe's total compilation times.}
    \label{fig:comptimes}
    \vspace{-2em}
\end{figure}

\revisiona{In Fig.~\ref{fig:comptimes} we present the distributions of DaCe's total compilation times on CPU, GPU, and FPGA for the set of benchmarks presented in Section~\ref{sec:eval-shmem}.
This time includes parsing the Python code, auto-optimizing and compiling with GCC, NVCC, Intel OpenCL SDK, or the Xilinx Vitis compiler.
90\% of the CPU and GPU codes compile in less than 15s, while there is a single outlier above one minute.
On FPGA architectures, synthesis, placement and routing typically takes hours, rendering DaCe's overhead negligible.}

\subsection{Evaluation}
\label{sec:eval-shmem}

In the following, we show results for single node shared memory parallel programs created using data-centric Python for CPU, GPU and FPGA, and compare these with other frameworks: NumPy over the CPython interpreter, Numba, Pythran, and CuPy. We collect a set of existing Python codes from different scientific and HPC domains~\cite{pyfai, cfdpython, nbody, dace-omen, omen-physics, cythontutorial, numbatutorial, mandelbrot, crc, numba-histogram, stockham, lenet, resnet, gt4py, cosmo1, cosmo2}, as well as a NumPy version of Polybench~\cite{polybench} ported from the C benchmark.
\revision{R4,R5}{In this adaptation, we strive to express the algorithms of the original benchmark in a way that is natural to a Python programmer.
E.g., in \lstinline{gemm}, \lstinline{2mm}, and \lstinline{3mm}, the matrix-matrix product is implemented with \lstinline{@}, Python's dedicated operator for matrix multiplication~\cite{pep456}.}
All the data used are double-precision floating point numbers or 64-bit integers for CPU and GPU, while the FPGA tests use single-precision floating point numbers and 32-bit integers.

\subsubsection{Experimental Setup}
The CPU and GPU evaluations are performed on a machine running CentOS 8, with 1.5 TB of main memory, two Intel Xeon Gold 6130 CPUs (2x16 cores), and an NVIDIA V100 GPU (CUDA version 11.1) with 32 GB of RAM.
We use CPython 3.8.5 as part of an Anaconda 3 environment.
We test NumPy 1.19.2 with Intel MKL support, Numba 0.51.2 with Intel SVML support, the latest Pythran version from their GitHub repository~\cite{pythran-repo} (commit ID 09349c5), and CuPy 8.3.0.
For all frameworks that need a separate backend compiler, we use GCC 10.2.0, with all the performance flags suggested by the developers.
To put high-performance Python into the perspective of low-level C implementations, we also compare the applications adapted from Polybench with the original Polybench/C~\cite{polybench} benchmark, compiled with GCC and the Intel C Compiler with automatic parallelization enabled (\texttt{icc -O3 -march=native -mtune=native -parallel}). %We experimentally find that icc utilizes parallelism and MKL BLAS calls on the kernels.

We evaluate FPGA performance on two different boards from either major FPGA vendor; a Bittware 520N accelerator with an Intel Stratix 10 2800 GX FPGA and the Xilinx Alveo U250 accelerator board. Intel FPGA kernels are built with the Intel OpenCL SDK for FPGA and Quartus 20.3 targeting the \texttt{p520\_max\_sg280h} shell, and Xilinx kernels are built with the Vitis 2020.2 compiler targeting the \texttt{xilinx\_u250\_xdma\_201830\_2} shell.

For the DaCe Python versions, we annotate types and symbolic shapes on the decorated functions to enable AOT compilation and work with FPGAs. We \textbf{do not} annotate loops as \lstinline{dace.map}s and keep them in their original form, leaving parallelization for the automatic heuristics (Section~\ref{sec:autoopt}). 

We compare the performance of the different frameworks and compilers using runtime as our primary metric of execution.
Unless otherwise mentioned, we run each benchmark ten times and report the median runtime and 95\% nonparametric confidence interval (CI)~\cite{garth}.

\subsubsection{Benchmarking Results}

\begin{figure}[t]
    \centering
    \includegraphics[width=.95\columnwidth]{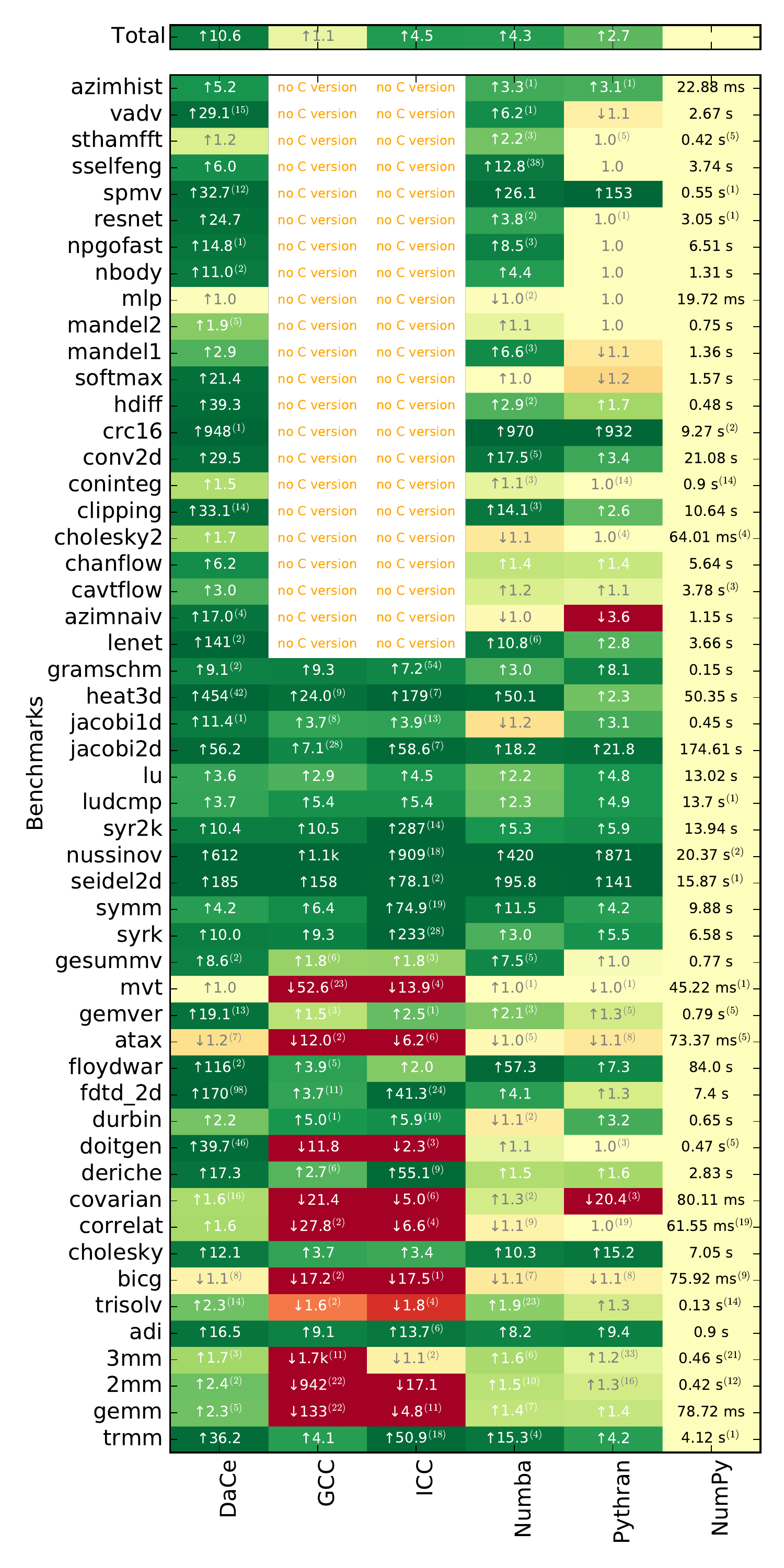}
    \vspace{-1em}
    \caption{CPU runtime and speedup over NumPy.}
    \label{fig:results-cpu}
\end{figure}

\begin{figure}[t]
    \centering
    \includegraphics[width=.89\columnwidth]{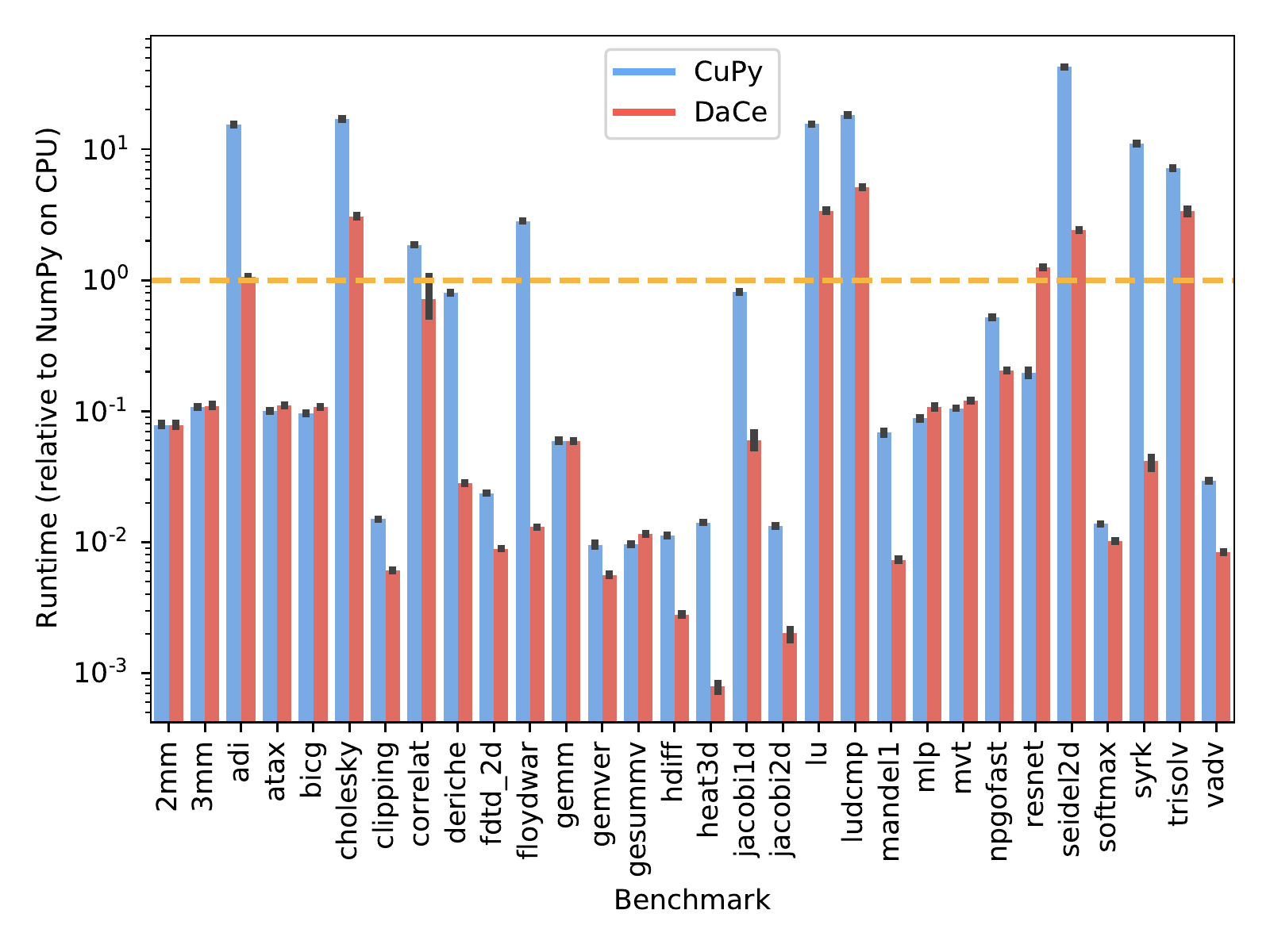}
    \vspace{-1em}
    \caption{CuPy and DaCe GPU runtime (lower is better).}
    \label{fig:results-gpu}
\end{figure}

CPU results are presented in Fig.~\ref{fig:results-cpu}.
The right-most column contains NumPy's execution runtime for each of the benchmarks annotated on the y-axis.
Each of the other columns contains the speedup (green tint and upward arrow) or slowdown (red tint and downward arrow) of execution compared to NumPy for each of the competing Python frameworks and Polybench C versions (compiled with ICC or GCC).
Furthermore, we compute the 95\% CI using bootstrapping~\cite{efron1992bootstrap} and annotate its size (as superscript in brackets) as a percentage of the median; values less than 1\% are omitted \revisionc{(Fig.~\ref{fig:results-cpu} uses vector graphics, and readers can zoom in with a PDF viewer if any values are not readable due to their size)}.
The upper part of the chart aggregates the benchmarks using the geometric mean of the speedups over NumPy.
Numba and Pythran have fallback modes for Python code that fails to parse.
In such cases, we measure the runtime of the fallback mode.

The figure shows an overall improvement in DaCe Python's performance, both over the Python compilers and the optimizing C compilers.
Specifically, the subgraph fusion transformation capabilities surpass those of Numba. This is especially apparent in stencils, where the difference can be in orders of magnitude. In applications such as \lstinline{crc16}, all compiled implementations successfully eliminate interpreter overhead.
Shorter kernels benefit from the C versions due to runtime (and timing) overhead mitigation. It also appears that with control-flow heavy codes (e.g., \lstinline{nussinov}), simple C code can be better optimized by GCC and ICC over generated code. It is worth noting that on some applications, NumPy is faster than the C versions: this is because of the performance benefits of vectorized NumPy code compared with explicit, sequential loops.
\revision{R4,R5}{An exceptional case is \lstinline{3mm}, which consists of three matrix multiplications.
ICC pattern matches the matrix-matrix product and links to MKL, achieving similar performance to the Python frameworks.
On the other hand, GCC does not compile the unoptimized C code to an executable that uses MKL, leading to lower performance.}

Fig.~\ref{fig:results-gpu} presents the runtime of applications that were successfully transformed to run on GPU. As with CPU, auto-optimizing DaCe consistently outperforms or is equivalent to CuPy, 3.75x (geomean) faster. The auto-optimization passes contribute to these results, mainly attributing to subgraph fusion and avoiding intermediate allocations on shorter applications. Due to redundant copy removal and view semantics being native to the SDFG, we see a particular improvement on stencils, e.g., \lstinline{heat3d}. 
\revisione{Although CuPy-optimized code could potentially employ similar transformations~\cite{cupy-fuse}, as far as we know, this cannot be performed out-of-the-box.
The user must explicitly define element-wise or reduction based kernels, significantly changing the code.}
There is one instance where CuPy outperforms DaCe --- \lstinline{resnet}. This is due to a suboptimal vectorized representation of convolution in the Python source code, which translates to a loop of summations. In our generated code, this automatically results in many unnecessary atomic operations, even if tiled. The issue can be easily mitigated with further manual transformations (changing the maps' schedules) after the fact.%\htor{we should explain why we suddenly subset here (carefully)}

FPGA results can be seen in Fig.~\ref{fig:results-fpga}, where there is no comparison point as no other framework compiles high-performance Python directly. Although both platforms use different languages and features (e.g., accumulators), applications can be synthesized for both from the same annotated Python code. There is a noticeable difference in performance, especially on stencil-like applications, likely resulting from Intel FPGA's compiler toolchain superior stencil pattern detection. However, this can also be mitigated with subsequent manual transformations on the SDFG or augmenting the automatic heuristics decision-making process to transform stencils explicitly.
\revisione{Library Node expansions take device-specific features into account. For example, when there are accumulations (e.g., GEMV), we take advantage of hardened 32-bit floating-point accumulation on Intel FPGAs, which allows single-precision numbers to be directly summed into an output register. On Xilinx FPGAs, and, in general, for 64-bit floating-point accumulation, such native support does not exist. Therefore, we perform accumulation interleaving~\cite{hls_transformations} across multiple registers to avoid loop-carried dependencies.}

\subsubsection{Discussion}
While DaCe already outperforms existing libraries, this is not the end of the optimization process. Instead, the generated SDFGs can be \textit{starting} points for optimization by performance engineers. The provided transformations and API can be used to eliminate sources of slowdown in the above applications or extended to use new, domain-specific transformations~\cite{andrei,licht2021stencilflow}. Furthermore, the applications can also be adapted to distributed memory environments, where the productivity and performance benefits can be even greater.

\section{Scalable Distributed Python}
\label{sec:scale}

The data-centric representation of Python programs can serve as a starting point for creating distributed versions.
\revision{R2,R5}{These distributed SDFGs abandon the global view of the data movement in favor of a local one. Like Message Passing, the flow of data in distributed memory is explicitly defined through Library Nodes.
This approach allows for fine-grained control of the communication scheme and better mapping of SDFGs to code using optimized communication libraries.}

In this Section, We show how to design transformations that specialize parallel map scopes to support distributed memory systems. We then show how to optimize such distributed data-centric programs. Finally, we show how developers can take control of the distribution entirely by expanding the original Python code with distributed communication while still allowing data-centric optimization to occur.

\subsection{Transforming for Scale} \label{sec:implicit-dist}
Leveraging the data-centric representation, we can create transformations that convert specific shared-memory parallel kernels into distributed memory.
The advantage of this approach is that once such a transformation is available, we can apply it to any subgraph in any SDFG that matches the same pattern. 
Furthermore, transformations can be compounded, building on each other to achieve complex results. 
We focus again on the \lstinline{gemm} kernel for illustrating the transformations. As we shall show, the transformations extend beyond and automatically distribute other kernels as well. %and the results below show how other Polybench kernels \textit{2mm}, \textit{3mm}, and others, all profit from it.

\begin{figure}[ht]
    \centering
    \includegraphics[page=6, width=.95\columnwidth]{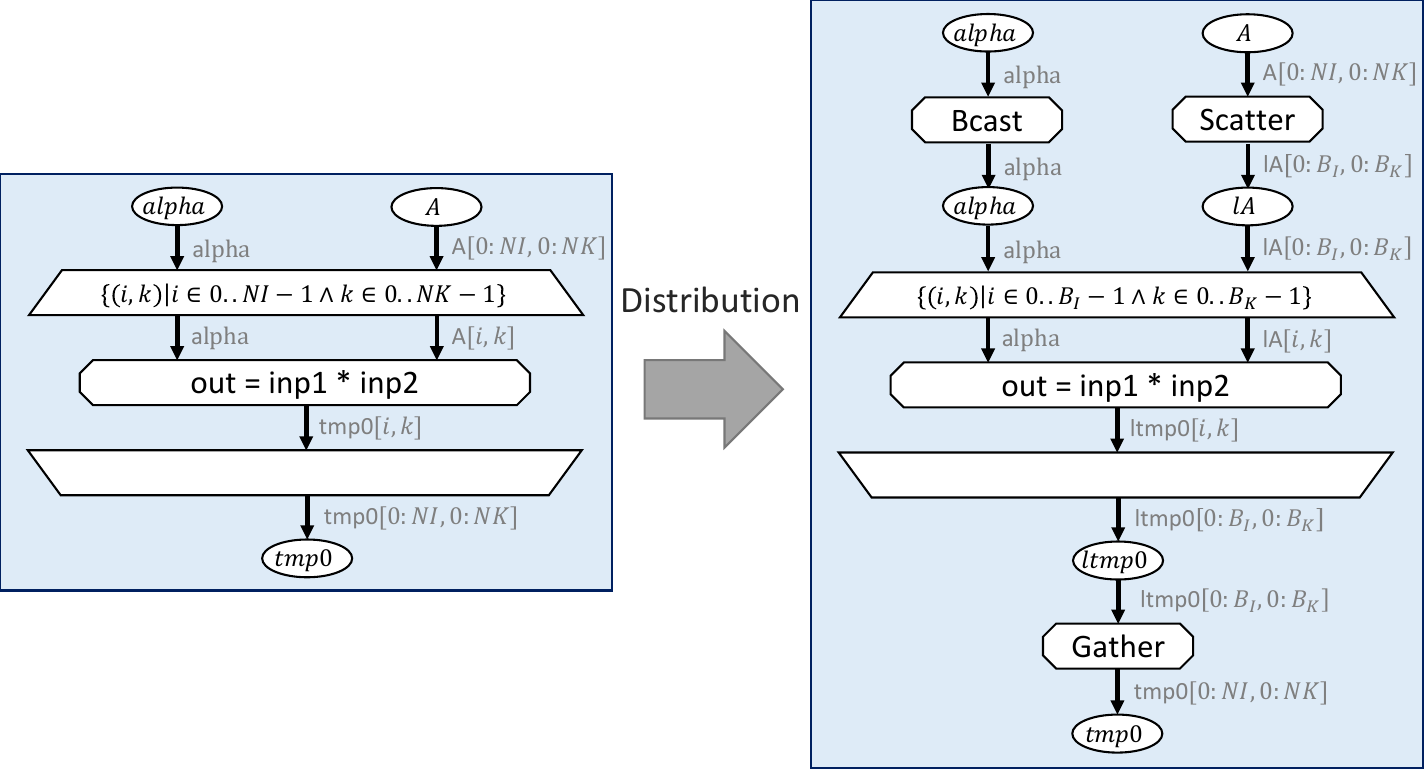}
    \vspace{-1em}
    \caption{Distribution of element-wise array operation.}
    \label{fig:gemm-distr-elwise-ops}
\end{figure}

\paragraph{Distributing global view element-wise operations}
We distribute these by following a scatter-gather pattern, broadcasting (scalars) or scattering (arrays) input data containers from the root rank to the machine nodes, performing local computation, and gathering or reducing the outputs.
By the nature of element-wise operations, careful selection of the array distribution parameters is not always necessary.
The primary constraint is that each rank receives matching subsets of data, allowing it to perform local computation without further communication.
Therefore, the most efficient distribution for contiguous arrays is to treat them as uni-dimensional and scatter them with \lstinline{MPI_Scatter} (1-D block distribution).
However, in cases where the result of an element-wise operation is consumed by, e.g., a matrix-matrix product or a stencil computation, it is beneficial to preserve the dimensionality of the arrays.
For this reason, the transformation has optional parameters for the block sizes per dimension, leading to \textit{block} or \textit{block-cyclic} distributions.
We offer implementations that use PBLAS~\cite{pblas} methods, such as \lstinline{p?gemr2d} and \lstinline{p?tran}, and MPI derived data types, which have previously demonstrated performance benefits~\cite{ddt}.
Applying the above transformation with \textit{block} distributions on the operation \lstinline{tmp0 = alpha * A} transforms the SDFG subgraph as shown in Fig.~\ref{fig:gemm-distr-elwise-ops}.
We emphasize that these transformations are applied to an SDFG \textbf{without} changing the data-centric program code:
\begin{lstlisting}[backgroundcolor=\color{peyellow}]
sdfg.apply(DistributeElementWiseArrayOp)
\end{lstlisting}
\begin{figure}[t]
    \centering
    \includegraphics[width=\columnwidth]{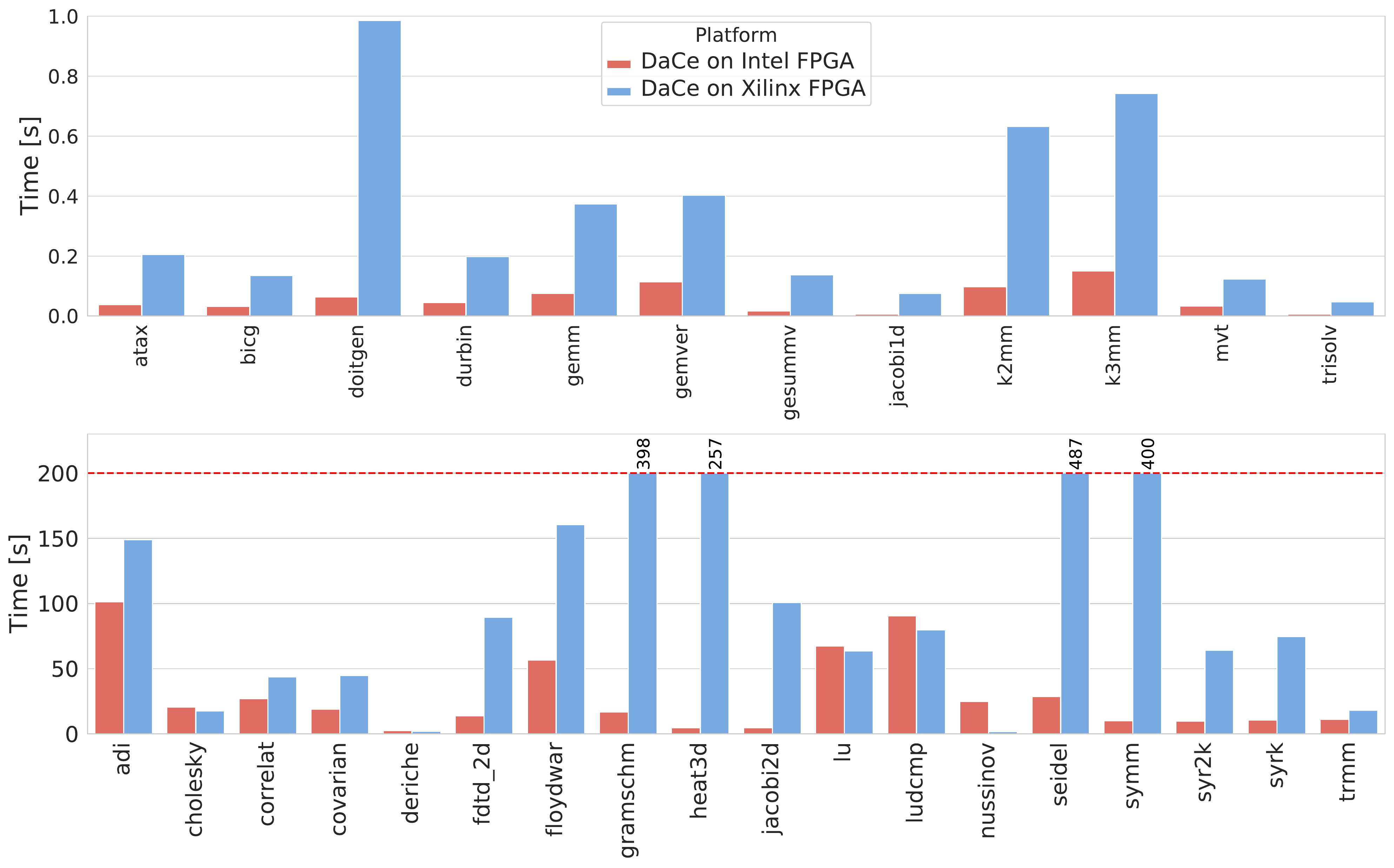}
    \vspace{-1em}
    \caption{FPGA runtime, \texttt{Large} instance, single precision.}
    \label{fig:results-fpga}
\end{figure}

\paragraph{Distributing Library Nodes}
We also create expansions for Library Nodes to distributed SDFG subgraphs.
For example, the matrix-matrix and matrix-vector products expand to the aforementioned PBLAS library calls, along with the corresponding distribution of inputs and gathering outputs.
\revisione{Using PBLAS requires the definition of a process grid.
The DaCe PBLAS library environment handles this automatically using BLACS~\cite{blacs}.
The grid's parameters are free symbols that can be chosen by the user or take default values.}

\subsection{Optimizing Communication} 
Creating distributed versions of the operations separately from each other will perform correctly but poorly on real applications. Thus, we can use the data-centric aspect of the SDFG IR, which can track access sets through memlets to remove such communication bottlenecks automatically. Doing so separately from the distribution transformations allows users to write more pattern-matching distribution transformations without worrying about inter-operation communication, on the one hand; and on the other hand, allows the system to find such optimizations in any input code (e.g., if manually-written with communication redundancy).

One example of such a transformation is redundant gather-scatter removal.
Consider the SDFG representation of \lstinline{gemm}, shown in Fig.~\ref{fig:gemm-strict}.
We distribute all three element-wise array operations and we expand the \lstinline{MatMult} node to a call to \lstinline{pdgemm}.
Due to the produced scatter and gather operations, the outputs \lstinline{tmp1} of \lstinline{pdgemm} and \lstinline{tmp2} of \lstinline{beta * C} will be connected to the last element-wise array operation \lstinline{tmp1 + tmp2} as shown in the following \textbf{pseudo-code}:
\begin{lstlisting}
pdgemm(..., ltmp1, ...)
tmp1 = gather(ltmp1)
ltmp1 = scatter(tmp1)
ltmp2 = beta * lC
tmp2 = gather(ltmp2)
ltmp2 = scatter(tmp2)
lC = ltmp1 + ltmp2
...
\end{lstlisting}
The above sequence of operations yields redundant communication on \lstinline{tmp1} and \lstinline{tmp2} and can therefore be omitted (the transformation for \lstinline{tmp1} is shown in Fig.~\ref{fig:gemm-distr-redundant-comm}).
By following the data movement and inspecting other Access nodes in the state, data dependencies of the global array can be inferred.
Furthermore, since we know the \lstinline{Scatter} and \lstinline{Gather} node semantics, we can check whether the data distributions match.
\revisione{We note that users can use the DaCe API to define transformations to, e.g., optimize the re-distribution of data when the distributions do not match.
If the distributions are 2D block-cyclic, such a transformation could, among other solutions, utilize PBLAS and a \lstinline{p?gemr2d} Library Node to efficiently bypass the \lstinline{Scatter} and \lstinline{Gather} operations.}

Combing the above transformations, the shared-memory \lstinline{gemm} program from Section~\ref{sec:py2dc} can be converted to distributed-memory as follows, again without altering the code of the dace-centric Python program (type annotations omitted for brevity):
\begin{lstlisting}
@dace.program
def gemm(alpha, beta, C, A, B):
    C[:] = alpha * A @ B + beta * C
\end{lstlisting}
\begin{lstlisting}[backgroundcolor=\color{peyellow}]
dist_sdfg = gemm.to_sdfg()
dist_sdfg.apply(DistributeElementWiseArrayOp)
dist_sdfg.expand_library_nodes('PBLAS')
dist_sdfg.apply(RemoveRedundantComm)
\end{lstlisting}

\subsection{Assuming Direct Control via Local Views}
\label{sec:explicit-distr}

The implicit, global view approach works well for a plethora of Python programs that make heavy use of high-level array operations.
However, as highly-tuned HPC applications often use specific partitioning schemes, our data-centric toolbox also provides explicit control via Python annotations. As opposed to the existing tools that manage communication implicitly, the aim of the interface is to use \textit{(num)pythonic} concepts to retain productivity while maximizing performance.
E.g., the \lstinline{jacobi_2d} stencil below would yield unnecessary \lstinline{Scatter} and \lstinline{Gather} collectives at every timestep:

\begin{figure}[t]
    \centering
    \includegraphics[page=8, width=.95\columnwidth]{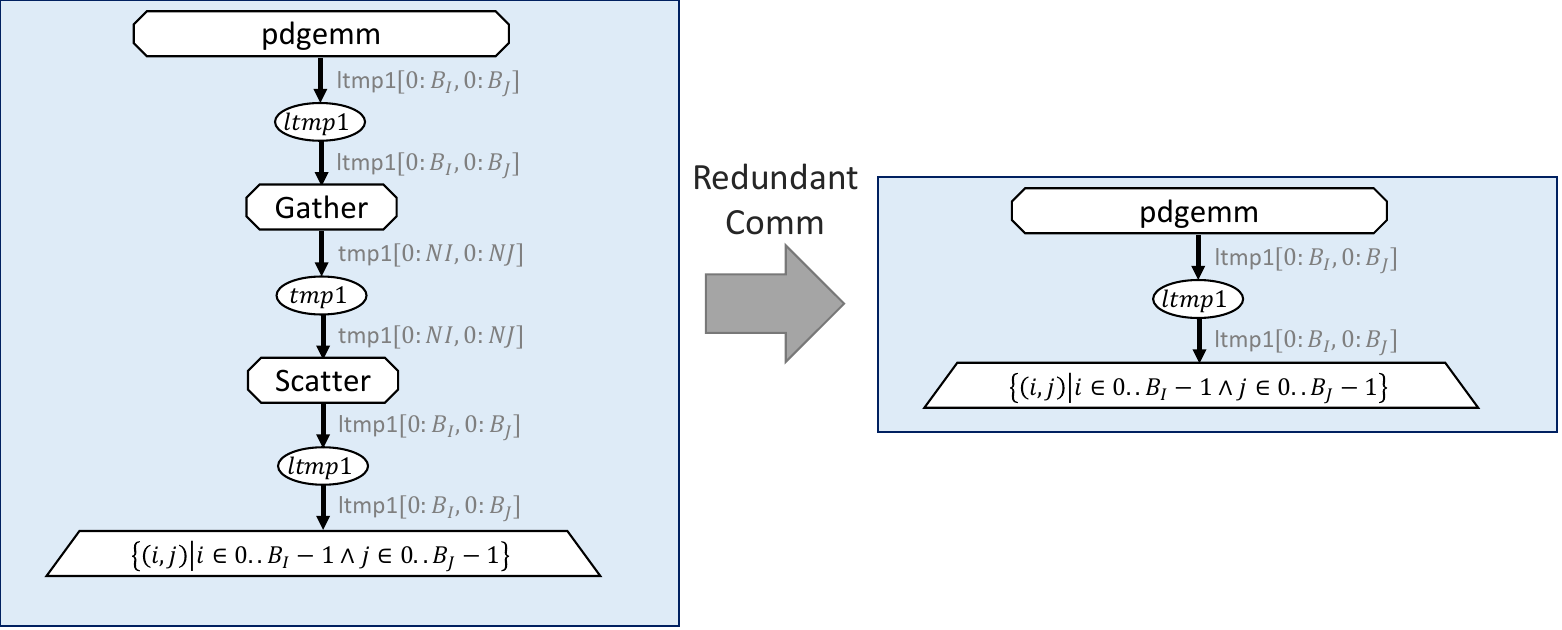}
    \vspace{-1em}
    \caption{Redundant communication elimination.}
    \label{fig:gemm-distr-redundant-comm}
\end{figure}
\begin{figure*}[t]
    \centering
    \includegraphics[width=.95\textwidth]{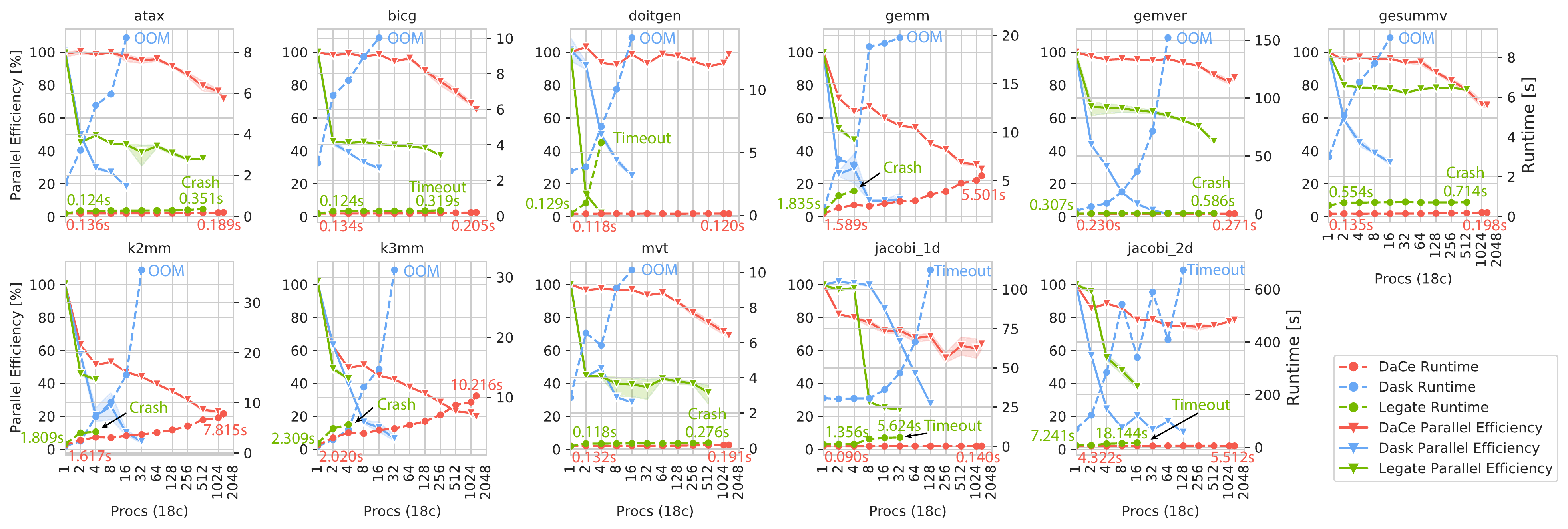}
    \vspace{-1em}
    \caption{\revisiondnotag{Distributed runtime (dashed lines) and scaling efficiency (solid lines) on 23,328 cores of Piz Daint.}}
    \label{fig:results-distr}
\end{figure*}

\begin{lstlisting}[basicstyle=\footnotesize\ttfamily]
@dace.program
def jacobi_2d(TSTEPS: dace.int32, A: dace.float64[N,N],
              B: dace.float64[N,N]):
    for t in range(1, TSTEPS):
        B[1:-1,1:-1] = 0.2 * (A[1:-1,1:-1] + A[1:-1,:-2]
            + A[1:-1,2:] + A[2:,1:-1] + A[:-2,1:-1])
        A[1:-1,1:-1] = 0.2 * (B[1:-1,1:-1] + B[1:-1,:-2]
            + B[1:-1,2:] + B[2:,1:-1] + B[:-2,1:-1])
\end{lstlisting}

For this reason, our data-centric approach allows the user to express arbitrary communication patterns by integrating explicit communication directly into Python. %\htor{to me, this looks no different than the transformations in 4.2 - should we make the difference more explicit (use different BG colors?) The call names seem to be very similar ... may need some elaboration as this is the second time here}
The above shared-memory data-centric Python program can be \textit{modified} to be distributed:

\begin{lstlisting}[basicstyle=\footnotesize\ttfamily]
@dace.program
def half_step(inpbuf: dace.float64[lNx+2, lNy+2],
              outbuf: dace.float64[lNx+2, lNy+2]):
  req = np.empty((8,), dtype=MPI_Request)
  dace.comm.Isend(inpbuf[1, 1:-1], nn, 0, req[0])
  # ...
  dace.comm.Irecv(inpbuf[1:-1, -1], ne, 2, req[7])
  dace.comm.Waitall(req)
  outbuf[1+noff:-1-soff, 1+woff:-1-eoff] = 0.2*(
      inpbuf[1+noff:-1-soff, 1+woff:-1-eoff] +
      # ...
      inpbuf[noff:-2-soff, 1+woff:-1-eoff])

@dace.program
def j2d_dist(TSTEPS: dace.int32, A: dace.float64[N, N],
             B: dace.float64[N, N]):
  lA = np.zeros((lNx+2, lNy+2), dtype=A.dtype)
  lB = np.zeros((lNx+2, lNy+2), dtype=B.dtype)
  lA[1:-1, 1:-1] = dace.comm.BlockScatter(A)
  lB[1:-1, 1:-1] = dace.comm.BlockScatter(B)
  for t in range(1, TSTEPS):
    half_step(lA, lB)
    half_step(lB, lA)
  A[:] = dace.comm.BlockGather(lA[1:-1, 1:-1])
  B[:] = dace.comm.BlockGather(lB[1:-1, 1:-1])
\end{lstlisting}

In the above program, we distribute the arrays \lstinline{A} and \lstinline{B} into 2D blocks at the beginning.
\revisione{The local views \lstinline{lA,lB} are then computed and communicated via explicit halo exchange, using \lstinline{Isend}, \lstinline{Irecv}, and \lstinline{Waitall} MPI calls in every time-step.}

While this approach is similar to the \textit{mpi4py} bindings, there are two distinct advantages to the data-centric approach with an explicit local view.
First, the MPI calls are integrated into the program's dataflow with Library Nodes, enabling the above transformations and other automatic code generation features such as overlapping.
\revisione{Second, explicit \textit{Isend} and \textit{Irecv} calls communicate strided data using the MPI vector datatype, avoiding extraneous copies.}
The latter is also an example of using symbolic information on the graph to assert that high performance is attained --- our MPI derived data type creation code, which is created once for each data type, relies on the symbol values not changing over the run.
E.g., the initialization of the \lstinline{inpbuf[1:-1, 1]} data type:
\begin{lstlisting}[basicstyle=\footnotesize\ttfamily]
MPI_Datatype ntype;
MPI_Type_vector(lNx, 1, lNy+2, MPI_DOUBLE, &ntype);
MPI_Type_commit(&ntype);
\end{lstlisting}
would raise a performance warning in DaCe Python if the sizes may change at runtime. This avoids potential mistakes that even experienced performance engineers can make in large HPC codes.

\subsection{Evaluation}

To measure the performance of distributed data-centric programs, we conduct scaling experiments on the multi-core partition of the Piz Daint supercomputer.
Each node has two 18-core Intel E5-2698v3 CPUs and 64GB of memory.
The nodes are connected through a Cray Aries network using a Dragonfly topology.
We benchmark a subset of the Polybench kernels from Section~\ref{sec:eval-shmem}, which could be automatically transformed to use distributed memory (Section~\ref{sec:implicit-dist}); \lstinline{adi}, \lstinline{bicg}, \lstinline{doitgen}, \lstinline{gemm}, \lstinline{gemver}, \lstinline{gesummv}, \lstinline{k2mm}, \lstinline{k3mm}, \lstinline{mvt}, \lstinline{jacobi_1d}, and \lstinline{jacobi_2d}.
\revision{R1,R4,R5}{We compare this work with Dask~\cite{dask} v2.31 and Legate~\cite{legate} (commit ID febd3bf~\cite{legate-repo}), two state-of-the-art distributed tasking Python frameworks, in weak scaling from 1 to 1,296 processes (648 nodes).
Dask Array~\cite{dask-array} scales a variety of NumPy workflows, including element-wise array operations, reductions, matrix-matrix products, and linear algebra solvers, among others.} 
Legate is providing a drop-in replacement for the NumPy API to accelerate and distribute Python codes.

\newcolumntype{P}[1]{>{\centering\arraybackslash}p{#1}}
\begin{table}[t]
\centering
\small
\renewcommand{\arraystretch}{0.8} 
\resizebox{\columnwidth}{!}{ % just to be sure that we don't overflow
\begin{tabular}{P{1.9cm}P{1.3cm}p{3.4cm}P{0.8cm}}
\toprule
\textbf{Benchmark}              & \textbf{F} &\textbf{Initial Problem Size} & \textbf{S.F.} \\ 
\midrule
\texttt{atax}                   & DaCe/Legate  & $20000, 25000$                    &  \multirow{2}{*}{all $\sqrt{S}$}  \\
{\selectfont\tiny$(M, N)$}      & Dask  & $10000, 12500$                    &  \\
\midrule
\texttt{bicg}                   & DaCe/Legate  & $25000, 20000$ & \multirow{2}{*}{all $\sqrt{S}$}	\\
{\selectfont\tiny$(M, N)$}      & Dask  & $12500, 10000$ &              \\
\midrule
\texttt{doitgen}                & DaCe/Legate  & $128, 512, 512$ & \multirow{2}{*}{$(S, -, -)$}	\\
{\selectfont\tiny$(NR, NQ, NP)$}& Dask  & $128, 512, 512$ &              \\
\midrule
\texttt{gemm}                	& DaCe/Legate  & $8000, 9200, 5200$ & \multirow{2}{*}{all $\sqrt[3]{S}$}	\\
{\selectfont\tiny$(NI, NJ, NK)$}& Dask  & $4000, 4600, 2600$ &    \\
\midrule
\texttt{gemver}                 & DaCe/Legate  & $10000$                    & \multirow{2}{*}{$\sqrt{S}$}  \\
{\selectfont\tiny$(N)$}      	& Dask  & $5000$                    &  \\
\midrule
\texttt{gesummv}                 & DaCe/Legate  & $22400$                    & \multirow{2}{*}{$\sqrt{S}$}  \\
{\selectfont\tiny$(N)$}      	& Dask  & $11400$                    &  \\
\midrule
\texttt{jacobi\_1d}             & DaCe/Legate  & $1000, 24000$                      &  \multirow{2}{*}{$( -, S)$}\\
{\selectfont\tiny$(T, N)$}      & Dask  & $1000 , 24000$                     &  \\
\midrule
\texttt{jacobi\_2d}             & DaCe/Legate  & $1000, 1300$                      &  \multirow{2}{*}{$( -, \sqrt{S})$}\\
{\selectfont\tiny$(T, N)$}      & Dask  & $1000 , 1300$                     &  \\
\midrule
\texttt{k2mm}                   & DaCe/Legate  & $6400$, $7200$, $4400$, $4800$ & \multirow{2}{*}{all $\sqrt[3]{S}$} \\
{\selectfont\tiny$(NI, NJ, NK, NM)$} & Dask  & $3200$, $3600$, $2200$, $2400$ &  \\
\midrule
\texttt{k3mm}                   & DaCe/Legate  & $6400$, $7200$, $4000$,  $4400$, $4800$ & \multirow{2}{*}{all $\sqrt[3]{S}$} \\
{\selectfont\tiny$(NI, NJ, NK, NL, NM)$} & Dask  & $3200$, $3600$,  $2000$,  $2200$, $2400$ &  \\
\midrule
\texttt{mvt}                 & DaCe/Legate  & $22000$                    & \multirow{2}{*}{$\sqrt{S}$}  \\
{\selectfont\tiny$(N)$}      	& Dask  & $11000$                    &  \\
\bottomrule
\end{tabular}
}
\caption{Distributed benchmarks, initial problem sizes for the different frameworks (F), and scaling factors (S.F.) as a function of the number of processes $S$.}
\vspace{-3em}
\label{tab:bench-distr}
\end{table}

We select initial problem sizes that fit typical HPC workloads without having excessive runtime.
The kernels' scaling factors and the initial problem sizes for each framework are presented in Tab.~\ref{tab:bench-distr}.
For kernels with computational complexity ranging from $O(n)$ to $O(n^2)$, we target a runtime in the order of 100ms.
For kernels with higher complexity, we target a runtime in the order of 1s.
We note that with the problem sizes selected for benchmarking data-centric Python and Legate, Dask either runs out of memory or exhibits unstable performance.
Thus, we halved the problem sizes for Dask but still encountered out-of-memory errors at larger node counts.
\revisiond{Furthermore, several issues rendered testing Legate at scale difficult.
With the assistance of the developers, many of those problems were solved; however, others remained.
We could only run each benchmark up to a fraction of the total nodes available, either due to runtime errors or because a single execution did not finish within 10 minutes of allocated time.
We annotate Fig.~\ref{fig:results-distr} with these errors.}

We ignore the time needed for initializing and distributing data and only measure the main computation and communication time.
We run all frameworks using default parameters where possible, i.e., ``auto" as chunk size in Dask and block distributions (not block-cyclic, which would allow the user to fine-tune the block-sizes) on a 2D process grid for DaCe.
Furthermore, we spawn one process per socket (2 processes per node) and 18 threads per process (equal to the number of physical cores in a socket).
\revisiond{Legate is executed with parameters suggested by the developers: two NUMA domains per node, 28GB of memory and one CPU with 16 threads per domain.}

Fig.~\ref{fig:results-distr} presents the distributed runtime and scaling efficiency of DaCe, Dask, and Legate.
DaCe exhibits four different efficiency patterns.
In \lstinline{doitgen}, the workload is distributed in an embarrassingly parallel manner, and no communication is needed.
Therefore, the efficiency is close to perfect.
The kernels \lstinline{atax}, \lstinline{bicg}, \lstinline{gemver}, \lstinline{gesummv}, and \lstinline{mvt} compute matrix-vector products and scale very well until 64 processes, where the drop in efficiency becomes more pronounced, but remains above 60\% for all data points.
The matrix-matrix product kernels, \lstinline{gemm}, \lstinline{k2mm}, and \lstinline{k3mm}, exhibit lower efficiency, which is consistent with the expected behavior of MKL-ScaLAPACK~\cite{kwasniewski2019red}.
Finally, the stencil kernels' efficiency (\lstinline{jacobi_1d} and \lstinline{jacobi_2d}) falls between the last two categories of kernels.
\revisiond{On the other hand, Dask and Legate exhibit a sharp drop in efficiency in almost all kernels immediately from the second process.}
An exception is the \lstinline{jacobi_1d} kernel, where Dask has higher efficiency than DaCe up to 16 processes.
However, this is possible due to Dask being much slower than DaCe (over 30x on the same problem size), allowing for much higher communication-computation overlap.
\revisiond{On BLAS-heavy benchmarks, Legate matches the runtime of DaCe on a single CPU, whereas in others we observe slowdowns of 1.7--15$\times$. %on the benchmarks that ,
In the benchmarks that scale to a large number of nodes (\lstinline{atax}, \lstinline{bicg}, \lstinline{gemver}, \lstinline{gesummv}, and \lstinline{mvt}), Legate's efficiency, after the initial drop, remains constant.}

\revision{R1,R5}{DaCe uses MPI for communication and links to the optimized Cray implementation.
Legate is built on top of the Legion runtime~\cite{bauer2012legion}, which employs the GASNet library~\cite{gasnet}, a networking middleware implementing global-address space.
Finally, Dask uses TCP for inter-worker communication.
Although the different communication approaches cannot explain all performance discrepancies, there are the most significant factors in the overall picture.}

\section{Productivity}

\revisionc{
Python is already a very productive language, especially for domain scientists, due to its rich ecosystem described in Section~\ref{sec:intro}.
Data-Centric Python is Python with extensions that themselves are valid Python syntax.
For example, the \lstinline{@dace.program} decorator follows the PEP 318 standard~\cite{pep318}.
Therefore, Data-Centric Python essentially inherits the Python language's programming productivity.
Since performance is the most important metric in HPC, scientific applications must eventually be lowered to a representation that is amenable to low-level optimizations for the underlying architectures.
Traditionally, this translates to writing these applications in C, C++, and Fortran, among other device-specific languages.
Therefore, an HPC project must force the domain scientists to sacrifice productivity and work directly on the lower-level languages or maintain two different code-bases.
DaCe, and other frameworks that accelerate Python, increase HPC productivity by bridging the code that domain scientists want to write with the code that achieves high performance.}

\section{Related Work}

Approaches similar to our own targeting Python code have already been introduced and compared with in Sections~\ref{sec:intro},~\ref{sec:portable}, and~\ref{sec:scale}. In this section, we further discuss relevant frameworks, libraries, and approaches towards the three \textbf{P}s.%, both within the scope of the Python ecosystem and beyond.

\paragraph{Productivity}
The complexity of optimizing applications, combined with the repetitive nature of performance engineering for specific domains, has given rise to a wide variety of Domain-Specific Languages (DSLs)~\cite{dawn, regent, darkroom, claw} and embedded DSLs, particularly in Python~\cite{gt4py,devito}. In the latter category, a notable example is deep learning frameworks, which use Python's various capabilities to construct readable code that performs well. PyTorch~\cite{pytorch} uses object-oriented programming to construct deep neural networks as modules, relying on reflection to detect parameters and nonblocking calls for asynchronous execution to avoid interpreter overhead. TensorFlow~\cite{tensorflow} used Python's weak typing system to construct graphs from Python functions but has recently transitioned to ``eager'' execution to improve productivity, making codes more readable and simplifying debugging.

\paragraph{Portability}

In the past three decades, compilers have undergone a transition from all-pairs solutions (between source languages and hardware platforms) to funneling through Intermediate Representations (IR), on which they can perform language- and platform-agnostic optimization passes.
\revisiona{Although DaCe is currently developed in Python, many other research compilers are based on the LLVM~\cite{llvm} infrastructure and IR.
There is an ongoing movement in the compiler community towards Multi-Level IRs~\cite{mlir}, in which a multitude of IR \textit{dialects} can retain domain- and platform-specific information, in turn enabling domain-specific optimizations~\cite{gysi2020domainspecific}.
MLIR performs optimization passes on each dialect to compile programs, followed by lowering passes to subsequent dialects, down to hardware mapping.
This feature could be utilized to implement DaCe Library Nodes.}
The data-centric transformation API is also shared by languages such as Halide~\cite{halide}, which enables users to invoke schedule optimizations separately from program definition. %DaCe Library Nodes are comparable to MLIR, in the sense that they allow to express domain-specific optimizations. 

\paragraph{Performance portability}
Aimed at keeping a consistent ratio of performance to peak performance across hardware~\cite{perfport}, it is the core premise of several standards~\cite{openmp, openacc, opencl, sycl, dpcpp}.
In directive-based frameworks~\cite{openmp, openacc}, \textit{pragma} statements are added to C/C++ and FORTRAN programs to introduce parallelism, similarly to our proposed annotations.
In kernel-based frameworks~\cite{opencl, sycl, dpcpp}, \textit{kernels} are constructed as functions with a limited interface and offloaded to target devices, such as CPUs, GPUs, or FPGAs.
As each platform requires its own set of directives, kernel parameters, or sometimes kernel implementations, programs often contain multiple codes for each target.
To resolve such issues, HPC languages such as Chapel~\cite{chapel} and HPF~\cite{hpf} define high-level implicit abstractions used as parallel primitives. Also popular in the HPC world are performance-portable libraries, embedded within C++, notably Kokkos~\cite{kokkos}, RAJA~\cite{raja}, and Legion~\cite{bauer2012legion} (which powers Legate~\cite{legate}).
These allow integrating heterogeneous and distributed systems through task-based abstractions, data dependency analysis (e.g., Legion's logical regions), and common parallel patterns. Such patterns can also be found in NumPy, and the SDFG can be seen as a generalization of these graphs with symbolic data dependencies.

\section{Conclusion}

\paragraph{Discussion}
\revisionc{While distributed SDFGs forego the global view of data movement to facilitate the design of custom communication schemes, future work could explore the trade-offs between a more \textit{pythonic} approach to communication and extracting the best performance.
Moreover, improvements to existing transformations, e.g., Vectorization, and implementation of new ones could increase the out-of-the-box performance, reducing the need for manual optimization.}

We present Data-Centric Python --- a high-performance subset of Python with annotations that produces supercomputer-grade HPC codes.
Based on the SDFG intermediate representation, we show how a pipeline of static code analysis and dataflow transformations can take input NumPy code, leverage its vectorized nature, and map it efficiently to CPUs, GPUs, and FPGAs, outperforming current state-of-the-art approaches on each platform by at least 2.4x.

The resulting data-centric programs effectively eliminate the performance pitfalls of Python, including interpreter overheads and lack of dataflow semantics for library calls, the latter being crucial for running at scale. Many even outperform baselines written in C code.
A key feature of the data-centric toolbox is giving users explicit control when necessary, rather than making assumptions at the framework level.
We evaluate Data-Centric Python on a distributed environment and show that the parallel efficiency remains above 90\%, even on hundreds of nodes.
These promising results indicate that productive coding with Python can scale \textit{and} map to heterogeneous compute architectures, setting the once-scripting language at the same level as FORTRAN, C, and other HPC giants.

\section{Acknowledgements}
This project received funding from the European Research Council (ERC) under the European Union’s Horizon
2020 program (grant agreements DAPP No. 678880, EPiGRAM-HS No. 801039, and DEEP-SEA No.955606).
The Swiss National Science Foundation supports Tal Ben-Nun (Ambizione Project No. 185778).
The project was also sponsored by the Paderborn University, under the DaceML-FPGA project, and Xilinx, under the Xilinx Adaptive Compute Cluster (XACC) program.
The authors would like to thank Mark Klein and the Swiss National Supercomputing Centre (CSCS) for access and support of the computational resources.

\bibliographystyle{ACM-Reference-Format}
\bibliography{dace_python}

\end{document}